\magnification=1200
\input iopppt.modifie
\input epsf
\def\received#1{\insertspace 
     \parindent=\secindent\ifppt\textfonts\else\smallfonts\fi 
     \hang{#1}\rm } 
\def\appendix{\goodbreak\beforesecspace 
     \noindent\textfonts{\bf Appendix}\secspace} 
\def\figure#1{\global\advance\figno by 1\gdef\labeltype{\figlabel}%
   {\parindent=\secindent\smallfonts\hang 
    {\bf Figure \ifappendix\applett\fi\the\figno.} \rm #1\par}} 
\headline={\ifodd\pageno{\ifnum\pageno=\firstpage\titlehead
   \else\rrhead\fi}\else\lrhead\fi}

\def\rrhead{\textfonts\hskip\secindent\it 
    \shorttitle\hfill\rm L\folio} 
\def\lrhead{\textfonts\hbox to\secindent{\rm L\folio\hss}%
    \it\aunames\hss} 
\footline={\ifnum\pageno=\firstpage
\hfil\textfonts\rm L\folio\fi}   
\def\titlehead{\smallfonts J. Phys. A: Math. Gen. {\bf 24} (1991)
L1031--L1036\hfil} 

\firstpage=1031
\pageno=1031

\jnlstyle
\jl{1}
\overfullrule=0pt

\letter{Critical behaviour near multiple junctions and dirty surfaces
in the two--dimensional Ising model}[Letter to the Editor]

\author{Ferenc Igl\'oi\footnote{\dag}{Permanent address: Central Research
Institute for Physics, H--1525 Budapest, Hungary}, Lo\"\ii c Turban and Bertrand
Berche}[Letter to the Editor]
 
\address{Laboratoire de Physique du Solide\footnote{\ddag}{Laboratoire associ\'e
au CNRS no 155}, Universit\'e de Nancy I, BP 239 \hfil\break F--54506 Vand\oe
uvre l\'es Nancy Cedex, France}

\received{Received 7 June 1991}

\abs
We consider $m$ two--dimensional semi--infinite planes
of Ising spins joined together through
surface spins and study the critical behaviour near to the
junction. The $m\!=\!0$ limit of the model---according to the replica
trick---corresponds to the semi--infinite Ising model in the presence of a random
surface field ({\smallfonts RSFI}). Using conformal mapping, second--order
perturbation expansion around the weak-- and strong--coupled planes limits and
differential renormalization group, we show that the surface critical
behaviour of the {\smallfonts RSFI} model is described by Ising
critical exponents with logarithmic corrections to scaling, while at multiple
junctions ($m\!>\!2$) the transition is of first--order. There is a
spontaneous junction magnetization at the bulk critical point.
\endabs

\vglue.8cm

\pacs{68.35Rh; 75.40Cx; 64.60Fr}

\submitted
\date

The critical behaviour of systems near a plane where translational invariance
is broken, is of considerable recent interest~[1, 2]. The prototype of these
problems is represented by the critical phenomena at a free regular surface
(semi--infinite criticality), and in more complex problems the effect of a
perturbation (e.g. surface coupling enhancement, interfaces, defects, random
surface fields etc.) can be analysed by relevance--irrelevance type
criteria~[3--5]. Such a stability analysis, however, does not work for the
two--dimensional Ising model in the case of marginal perturbations caused by
a defect line~[3, 4] or a random surface
field ({\smallfonts RSFI})~[5], when the defect exponent $y_d\!=\!0$. In the
former case non--universal critical behaviour was found by an
exact calculation~[6, 7], while
for the {\smallfonts RSFI} model no definite answer is known yet. One of our purpose in
the present Letter is to clarify the critical behaviour of the {\smallfonts RSFI} model.
{\par\begingroup\parindent=0pt
\epsfxsize=9truecm
\topinsert
\centerline{\epsfbox{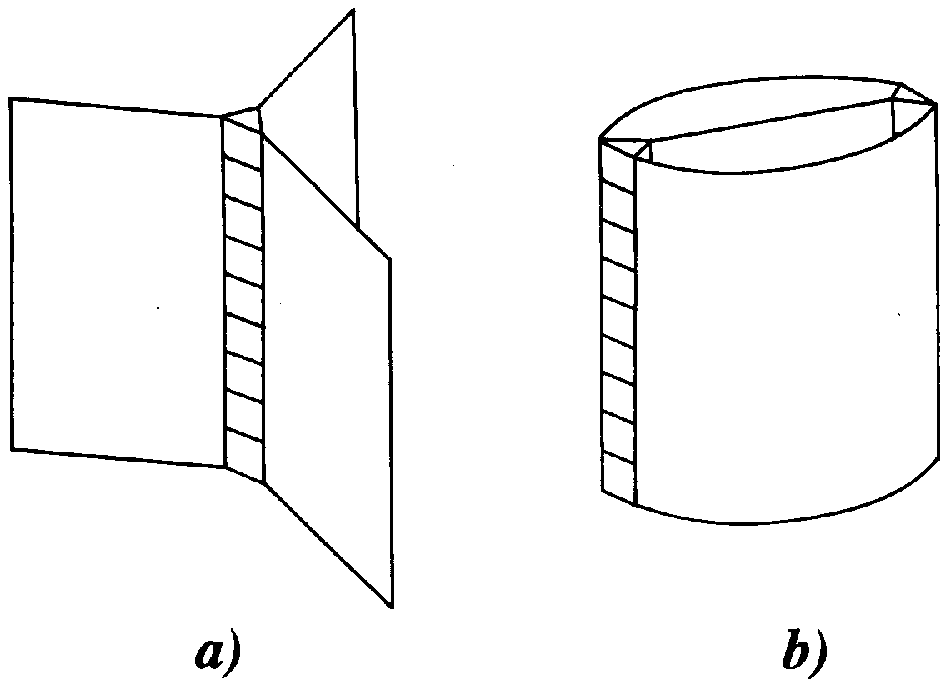}}
\smallskip
\figure{$(a)$ $m\!=\!3$ semi--infinite planes of Ising spins joined together by
surface spins and $(b)$ the corresponding system in the strip geometry.}  
\endinsert 
\endgroup
\par}
To study this problem we introduce a series of models consisting of $m$
semi--infinite planes of Ising spins where the spins at different surfaces
are joined together by nearest neighbour couplings (see figure~1($a$)).
(We note that in a recent paper Indekeu and Nikas~[8] introduced a junction as a
product of surface spins and studied the wetting phenomena in the $m\!=\!3$
system in the frame of Landau theory). The Hamiltonian of the system is given by:
$$ 
H=\sum_{p=1}^m H_p + V 
\eqno(1)
$$
where
$$
\fl-\beta H_p=\sum_{y=-\infty}^{\infty} \sum_{x=1}^{\infty}\left[
J_y \sigma^p(y,x) \sigma^p(y+1,x)+J_x \sigma^p(y,x) \sigma^p(y,x+1)\right] 
\eqno(2)
$$
and
$$
-\beta V=J \sum_{y=-\infty}^{\infty}\sum_{p<p'}
\sigma^p(y,1)\sigma^{p'}(y,1)
\eqno(3)
$$
Here $\sigma^p(y,x)=\pm 1$ are Ising spins at position $y,x$
on the $p$-th plane, and $\beta=1/k_B T$.

For $m\!=\!1$ and $m\!=\!2$ we obtain a semi--infinite system and two
semi--infinite systems joined by a defect
line, respectively, while for $m\!\geq\!3$  as a possible physical realization
one can imagine interacting magnetic ions segregated along various planar
grain boundaries, three of which meet along a linear junction. Due to pair
interactions in the junction (equation~(3)) the perturbation represented by $V$
is {\it marginal} in the ordinary surface transition point for all $m\!\ne\!1$;
thus to clarify the actual critical behaviour one needs detailed investigations.

Now we show that the multiple junction problem is connected to the {\smallfonts RSFI}
model defined by the Hamiltonian:
$$
-\beta \tilde H_p=-\beta H_p + \sum_{y=-\infty}^{\infty}h(y) \sigma^p(y,1)
\eqno(4)
$$
where the random surface field $h(y)$ has a Gaussian distribution:
$$
P[h(y)]={1 \over \sqrt{2\pi \Delta^2}}\exp \left(-{h^2(y) \over
2\Delta^2}\right)  
\eqno(5)
$$
Since the disorder is quenched, it is the free energy rather than the
partition function which must be averaged. Using the replica trick:
$\langle\log Z\rangle\!=\!\lim_{m\!\to\!0}\left[(\langle Z^m\rangle\!-\!1)/m
\right]$ one can easily show that the effective Hamiltonian of the problem is
just the  $m\!\to\!0$ limit of equations~(1)--(3) with $J\!=\!\Delta^2$.

In general we are interested in the critical behaviour near to the junction,
e.g. we look for the critical exponent $\eta_m$ describing the decay of
spin--spin correlations $\langle
\sigma^p(0,1)\sigma^p(y,1)\rangle\!\propto\!\vert y\vert^{-\eta_m}$ when the
system is at the bulk critical point. Now we suppose that critical correlations
in the model transform covariantly under a conformal transformation. It is
exactly known for $m\!=\!1$ and $m\!=\!2$~[9], on the other hand the
gap--exponent relation~(6) might be valid, even if the system is not conformally
invariant~[10]. In the following we map the system onto the strip
geometry~[11], where the calculation is usually simpler to perform.  Denoting
the points on the $p$th plane by the complex number $z_p$, then the conformal
transformation $w_p\!=\!(L/\pi)\log z_p,~p\!=\!1,2,...,m,$ maps the
semi--infinite planes onto strips of width $L$, and the surface spins at both
ends of the strips are connected to each other with the same type of coupling as
in the plane geometry equation~(3) (see figure 1(b)). 

The critical exponents in the strip geometry can be calculated from the
finite--size behaviour of the correlation length~[11]. More precisely we
consider the extreme--anisotropic limit~[12] of the model, when the transfer
matrix along the strip is expressed as $T\!=\!\exp(-a\hat H)$, where $a$ is the
lattice spacing and $\hat H$ is a quantum Hamiltonian. Then, following Cardy's
derivation~[13], one can show that the spectrum of the critical Hamiltonian
operator $\hat H$ in the large-$L$ limit is given as: 
$$
E_i-E_0={\pi\over L}v_sx_i
\eqno(6)
$$
where $E_0$ and $E_i$ are the ground state and the $i$th excited state of $\hat
H$, respectively, and $v_s$ is a normalizing factor, the so--called sound
velocity. The set of critical dimensions $x_i$ describes the decay of
correlations of scaling operators $\Phi^i$ along the junction in the plane
geometry: $\langle\Phi^i(0)\Phi^i(z)\rangle\!\propto\!\vert z\vert^{-2x_i}$. For
the spin operator we have $\eta_m\!=\!2x_s$. The spectrum in~(6) usually has a
tower--like structure; the levels in the same tower differ by an integer from
the lowest one: $x_i\!=\! x_i^0\!+\!l$; $l\!=\!0,1,2,\dots,$ and $x_i^0$ is the
critical dimension of a primary operator~[11].

For our model, the quantum Hamiltonian is given as
 
$$\eqalignno{
&\hat H=\sum_{p=1}^m \hat H_p + \hat V \eqno(7)\cr
&\hat H_p= -\sum_{x=1}^{L-1} \sigma_z^p(x) \sigma_z^p(x+1)-h\sum_{x=1}^{L}
\sigma_x^p(x)  \eqno(7a)\cr
&\hat V=-\lambda_1\sum_{p<p'}\sigma_z^p(1) \sigma_z^{p'}(1)
-\lambda_L\sum_{p<p'}\sigma_z^p(L) \sigma_z^{p'}(L)
\eqno(7b)}
$$
In (7$a$, $b$) $\sigma_x^p$ and $\sigma_z^p$ are Pauli matrices at chain $p$ on
site $x$, the bulk critical point corresponds to $h=1$~[12], while in the units
used in (7) $v_s\!=\!2$~[14]. Note that we put different values of the couplings
at both ends of the chains ($\lambda_1$,$\lambda_L$): the model in (1--3)
corresponds to $\lambda_1\!=\!\lambda_L\!=\!\lambda$.

We mention that one may also consider systems composed of planes having two
free surfaces, say at $x\!=\!0$ and $y\!=\!0$. For these '{\it half--infinite}'
systems, in the Hamiltonian equations~(2),~(3), the summations over $y$ run
from $0$ to $\infty$. Now in the transformed geometry the strips are coupled
only at one edge and the others are free; furthermore in the transformation
$\pi$ is replaced by $\pi/2$, the angle at the corner~[11]. In this case 
in~(7b) we have $\lambda_1\!=\!\lambda$ and $\lambda_L\!=\!0$.

To calculate the spectrum of~(7) one has to keep in mind the possible 
presence of strong logarithmic corrections to scaling, and therefore to try to
push analytical calculations as far as possible. Since for general value of~$m$
no exact solution can be found, we perform  a perturbation expansion around the
uncoupled chains limit and sum up the most diverging contributions by using
the differential renormalization group technique.

The actual calculation of the coefficients of the weak--junction expansion is
rather cumbersome; therefore here we present only the final result of the
second-order calculation; details of the derivation will be published
elsewhere~[15]. The first gap of the system---corresponding to magnetic
excitations---in the large-$L$ limit is given by:
$$
\fl E_1-E_0={2\pi \over L} \left[{1\over
2}-(m-1){1 \over \pi}(\lambda_1 +\lambda_L)+(m-1){2 \over \pi^2}\lambda_1
\lambda_L\right.\cr
\fl\qquad\qquad\left.-(m-1)(m-2)\left({2 \over \pi^2}\log L
-C\right)(\lambda_1^2 +\lambda_L^2)+...\right] + O(1/L^2) 
\eqno(8)
$$
where $C\!=\!0.03672...$ is a constant, and according to~(6) the quantity in the
square bracket is just the leading magnetic exponent~$x_s$. The main observation
concerning equation~(8) is that in first order $x_s$ is regular and coupling
dependent, but the second--order coefficient diverges as~$\log L$. The
perturbation series for higher gaps shows the same qualitative picture. 
On the basis of the differential renormalization group~[16] one assumes that
the higher--order terms of the expansion in~(8) are divergent too, but these
singular terms sum up to a regular contribution when the perturbation is
marginally irrelevant.

To show this we start writing the differential renormalization group equations of
the problem under a change of the length scale~$e^l$ in the
form~[16]:
$$\eqalign{
&{d\lambda_1 \over d l}=y_d\lambda_1
+b\lambda_1^2+O(\lambda_1^3)\cr 
&{d\lambda_L \over d l}=y_d\lambda_L
+b\lambda_L^2+O(\lambda_L^3)\cr} 
\eqno(9)
$$
where the defect exponent $y_d\!=\!0$ and the perturbation is marginally
irrelevant and marginally relevant for $b\!<\!0$ and $b\!>\!0$, respectively. The
solutions of (9) are given as $\lambda_1(l)\!=\!\lambda_1/(1\!-\! b\lambda_1 l)$
and $\lambda_L(l)\!=\!\lambda_L/(1\!-\! b\lambda_Ll)$, respectively. From the
transformation form of the inverse correlation length
$$
\xi^{-1}_n(\lambda_1,\lambda_L,L^{-1})=e^{-l}\xi^{-1}_n(\lambda_1(l),
\lambda_L(l),L^{-1}e^{l})
$$
one obtains the scaling prediction of the gap for finite systems:
$$
E_n-E_0=L^{-1} \Phi_n \left( {\lambda_1 \over 1-b\lambda_1 \log L},
{\lambda_L \over 1-b\lambda_L \log L} \right) 
\eqno(10)
$$
Expanding $\Phi_n$ up to second order in powers of $\lambda_1$, $\lambda_L$
one can verify that its form is compatible with the expansion in (8) with
$b\!=\!(2/\pi)(m\!-\!2)$. One can verify similarly that the scaling form (10)
is also valid for higher gaps with the same value of $b$. This fact is due to the
structure of the second--order {\it degenerate} perturbation calculation:
different gaps are represented by different secular matrices, but all the
matrix elements are the same function of $\lambda_1$, $\lambda_L$ and $L$,
independently of the matrix.

Thus we arrive at the conclusion, that the relevance--irrelevance behaviour
of the perturbation caused by the junction depends on the value of $m$: for
positive defect couplings it is marginally irrelevant and marginally relevant
for $m\!<\!2$ and $m\!>\!2$, respectively. In the borderline case $m\!=\!2$ the
perturbation is fully marginal, the critical exponents are coupling
dependent~[6, 7].

For $m\!<\!2$ in the large-$L$ limit: $\vert b\vert\lambda_1\log L\!\gg\!1$,
$\vert b\vert\lambda_L\log L\!\gg\!1$, the finite--size behaviour of the
magnetic exponent is given by: 
$$
x_s={1 \over 2} - { m-1 \over 2-m} {1 \over \log L}+...
\eqno(11)
$$ 
Concerning the surface transition of the {\smallfonts RSFI}
model ($m\!=\!0$) we obtain Ising critical exponents with logarithmic correction
to scaling 
$$
x_s^R={1 \over 2}\left(1+{1 \over \log L}+\dots\right)
\eqno(11a)
$$
which is one of the main results of our paper. The finite--size form of the
exponent $x_S^R$ in~(11a) gives a theoretical basis to deduce effective
exponents from {\smallfonts MC} simulations or from transfer matrix
calculations. We note, that similar behaviour---exponents of the pure model
with logarithmic corrections---has been observed for the magnetization and the
susceptibility of the random--bond two--dimensional Ising model~[17], which can
be considered as the bulk analogue of the {\smallfonts RSFI} model.

Now we turn to the problem of multiple junctions ($m\!>\!2$) with ferromagnetic
interactions. In this case the perturbation is {\it marginally~relevant},
consequently the critical behaviour is controlled by a {\it new fixed point}. We
believe that the system for non--zero defect couplings undergoes a first--order
transition, there is a spontaneous junction magnetization at the bulk critical
point. To prove this conjecture we consider the strong defect limit of the
problem and investigate the stability of the corresponding fixed point.

In the limit $\lambda_1\!\to\!\infty$ and/or $\lambda_L\!\to\!\infty$ all the
spins in the junctions are parallel, consequently the ground state of the system
is two--fold degenerate, i.e., the lowest gap is zero and there is a spontaneous
junction magnetization of the system. In this limit the spectrum of $\hat H$
can be constructed as the direct product of $m$ Virasoro algebras corresponding
to the spectrum of the Ising model at the extraordinary transition point~[11].

The stability of this fixed point is determined through the size dependence of
the lowest gap for finite defect couplings. In the {\it half--infinite} case,
$\lambda_1\!\gg\!1$, $\lambda_L\!=\!0$ in first order of $1/\lambda_1$ the gap is
given by~[14]
$$
E_1-E_0={m \over (m-1)2^{m-2}}{1 \over \lambda_1^{m-1}}\left( {1 \over
L} \right) ^{m/2}+\dots 
\eqno(12)
$$
while for non--zero $\lambda_L$ the leading finite--size dependence of the gap
remains the same with a more complicated prefactor than in~(12).

According to (12) for $m\!>\!2$ at finite couplings the first gap vanishes
faster than $1/L$, thus from (6) $x_s\!=\!0$ and the perturbation is irrelevant.
The ground state of the system is degenerate and there is a spontaneous junction
magnetization at the bulk critical point in accordance with our claim. We
note that similar mechanism has been observed for other problems
where surface or interface ordering take place at the bulk critical point:
the lowest gap vanishes algebraically with the size of the system~[18--20].
For $m\!=\!2$ according to~(12) the perturbation is
marginal, $x_s$ is coupling dependent, while for $m\!<\!2$ the gap vanishes
slower than $1/L$, thus the extraordinary transition is unstable in this region.

Finally we turn to discuss the critical behaviour near to a junction with
anti\-ferro\-magnetic~({\smallfonts AF}) interactions. In this case according 
to equations~(9) the signs of $b$ for marginally irrelevant and marginally
relevant perturbations are interchanged with respect to the ferromagnetic
junction. As a consequence for~$m\!>\!2$
the perturbation at the ordinary surface
transition point is {\it marginally~irrelevant}; the critical exponent~$x_s$
together with the logarithmic finite size correction is given by~(11).

It is interesting to study the strong junction limit for {\smallfonts AF}
couplings, which gives qualitatively different results for $m\!=$odd and
$m\!=$even. For $m\!=$odd, due to fru\-stra\-tion, there is no extra\-ordinary
tran\-si\-tion even for infinitely strong couplings in the defect: the first gap
is proportional to $L^{-1/2}$~[15]. This phenomenon resembles the absence of
phase transition at $T\!=\!0$ for super--frustrated models~[21]. On the other
hand for $m\!=$even there is an extra\-ordinary tran\-sition for infinitely
strong {\smallfonts AF} couplings. The perturbation to the first gap is marginal
in leading order: $E_1\!-\!E_0\!\propto\!1/(\lambda L)$, thus further analysis is
needed to decide on the type of transition in this case.

\ack FI is indebted to the Laboratoire de Physique du
Solide for hospitality. He is indebted to J Indekeu for valuable
discussions and sending the preprint in~[8] prior to publication. Useful
discussions with T W Burkhardt about conformal mapping of the problem are
also gratefully acknowledged.

\references

\numrefbk{[1]}{Binder K 1983}{Phase Transitions and Critical Phenomena}{vol 8,
ed C Domb and J L Lebowitz (London: Academic)}

\numrefbk{[2]}{Diehl H W J 1986}{Phase Transitions and Critical Phenomena}{vol 10,
ed C Domb and J L Lebowitz (London: Academic)}

\numrefjl{[3]}{Bray A J and Moore M A 1977}{\JPA}{10}{1927}

\numrefjl{[4]}{Burkhardt T W and Eisenriegler E 1981}{\PR\ {\rm B}}{24}{1236}

\numrefjl{[5]}{Diehl H W and N\"usser A 1990}{\ZP\ {\rm B}}{79}{69}

\numrefjl{[6]}{Bariev R Z 1979}{Zh. Eksp. Teor. Phys.}{77}{1217}

\numrefjl{[7]}{McCoy B M and Perk J H H 1980}{PRL}{44}{840}

\numrefjl{[8]}{Indekeu J and Nikas Y}{Preprint}{}{Leuven}

\numrefjl{[9]}{Turban L 1985}{\JPA}{18}{L325}

\numrefjl{ }{Henkel M, Patk\'os A and Schlottmann M 1989}{Nucl.
Phys.\ {\rm B}}{314}{609}

\numrefjl{[10]}{Guim I, Burkhardt T W and Xue T 1990}{\PR\ {\rm B}}{42}{10298}

\numrefbk{[11]}{Cardy J L 1987}{Phase Transitions and Critical Phenomena}{vol 11,
ed C Domb and J L Lebowitz (London: Academic)}

\numrefjl{[12]}{Kogut J 1979}{\RMP}{51}{659}

\numrefjl{[13]}{Cardy J L 1984}{\JPA}{17}{L385}

\numrefjl{[14]}{Burkhardt T W and Guim I 1984}{\JPA}{17}{L385}

\numrefjl{[15]}{Igl\'oi F, Turban L and Berche B, unpublished}{}{}{}

\numrefjl{[16]}{Cardy J L 1986}{\JPA}{19}{L1093}

\numrefjl{[17]}{Wang J S, Selke W, Dotsenko Vl S and Andreichenko V B
1990}{Europhys. Lett.}{11}{301}

\numrefjl{[18]}{Burkhardt T W and Igl\'oi F 1990}{\JPA}{23}{L633}

\numrefjl{[19]}{Igl\'oi F, Berche B and Turban L 1990}{\PRL}{65}{1773}

\numrefjl{[20]}{Igl\'oi F 1990}{\PRL}{64}{3035}

\numrefjl{[21]}{S\"ut\H o A 1981}{\ZP}{44}{121}

\vfill \eject
\bye